\documentclass[pra,twocolumn,tightenlines,showpacs,preprintnumbers,amsmath,amssymb]{revtex4}

\usepackage{graphicx}
\usepackage{dcolumn}
\usepackage{bm}
\usepackage{color}

\newcommand{\fref}[1]{Fig.~\ref{#1}}

\newcommand{\cref}[1]{chapter~\ref{#1}}

\newcommand{\Cref}[1]{Chapter~\ref{#1}}

\usepackage{ulem}  
\begin{document}

\preprint{}
\input{epsf.tex}

\epsfverbosetrue

\title{Van der Waals Interactions among Alkali Rydberg Atoms with Excitonic States}

\author{Hashem Zoubi}
\email{hashem.zoubi@aei.mpg.de}
\affiliation{Institute for Theoretical Physics, Institute for Gravitational
  Physics (Albert Einstein Institute), Leibniz University Hannover,
  Callinstrasse 38, 30167 Hannover, Germany}

\date{08 May, 2015}

\begin{abstract}
We investigate the influence of the appearance of excitonic states on van der
Waals interactions among two Rydberg atoms. The
atoms are assumed to be in different
Rydberg states, e.g., in the $|ns\rangle$ and $|np\rangle$ states. The resonant dipole-dipole interactions yield symmetric
and antisymmetric excitons, with energy
splitting that give rise to new resonances as the atoms approach each other. Only
far from these resonances the van der Waals coefficients, $C_6^{sp}$, can be
defined. We calculate the $C_6$ coefficients for alkali atoms and present the
results for lithium by applying perturbation theory. At short
interatomic distances of several $\mu m$, we show that the widely used simple model of two-level systems
for excitons in Rydberg atoms breaks down, and the correct representation
implies multi-level atoms. Even though, at larger distances one can keep the two-level systems but in including van
der Waals interactions among the atoms.
\end{abstract}



\maketitle

\section{Introduction}

Rydberg atoms, i.e.~atoms with large principal quantum numbers $n\gg 1$, have
been the subject of extensive research due to their unique optical and
electrical features \cite{Gallagher1994}. They are an ideal platform for the
study of a wide range of quantum phenomena, mainly since their properties
scale strongly with $n$, and often turning them extraordinary relative to
those of ground-state atoms. The property which has stimulated the most recent
experimental and theoretical work is strong interactions among Rydberg atoms that are separated by relatively large distances of several $\mu m$ \cite{Saffman2010,Low2012,Jaksch2000,Lukin2001,Pohl2010,Kiffner2013}.

On the other hand, electronic excitations can be delocalized among distant atoms through resonant dipole-dipole interactions to form collective
electronic excitations that are termed excitons
\cite{Davydov1971,Agranovich2009}. Different processes of resonant energy transfer
 are possible \cite{Gallagher2008}, here we concentrate on the exchange type,
 e.g. $ns\ +\ np\ \leftrightarrow\ np\ +\ ns$. Excitons have been introduced to a
 system of ultracold atoms in an optical lattice with lowest excited states
 where van der Waals interactions are negligible \cite{Zoubi2007,Zoubi2013}. The formation of excitons in a cluster of Rydberg atoms
has been investigated, but van der Waals forces were completely neglected and the discussion
limited to two-level atoms \cite{Wuster2010,Mobius2011,Ates2008}. In previous work we investigated
the influence of van der Waals interactions on the formation of
excitons in an aggregate of two-level Rydberg atoms \cite{Zoubi2014}. Coherent energy transfer among Rydberg atoms
that induces by resonant dipole-dipole interaction has been realized
experimentally at large interatomic distances of tenths $\mu m$ where van der Waals
forces are negligible \cite{Browaeys2014,Adams2015}.

An excitonic state contains at least two different
atomic states. The simplest case is of two atoms in which one in the
$|ns\rangle$ state and the other in the $|np\rangle$ state. The discussion can
be limited to dipole-dipole interactions, where dipole-quadrupole, quadrupole-quadrupole and higher order
interactions are neglected. Approximate long-range potentials can be derived
by applying perturbation theory up to the second order in the dipole-dipole
interactions. The lowest order term of the perturbation series results in
resonant dipole-dipole interaction of the form $C_3^{sp}/R^3$, where
$R$ is the interatomic distance with the resonant dipole-dipole coefficient
$C_3^{sp}$. This interaction leads
to a coherent mixing of the two possible states, which are $|ns,np\rangle$ and
$|np,ns\rangle$. The second order term is of the van
der Waals type of the form $C_6^{sp}/R^6$, with the van der Waals
coefficient $C_6^{sp}$. As was shown in our previous work \cite{Zoubi2014}, van
der Waals interactions result in energy shifts that significantly influence the formation of excitons when atoms approach
each other. Higher order terms play significant roles and can change completely
the long-range interaction potentials \cite{Boisseau2002,Porsev2014}. Furthermore, the appearance of resonances breaks down the validity of the
perturbative calculation and then other techniques are required,
e.g. in using direct Hamiltonian diagonalization \cite{Schwettmann2006}. Moreover, non-adiabatic interactions
between Rydberg atoms in different electronic surfaces can be important
\cite{Robicheaux2004}.

Long-range van der Waals interactions among pairs of Rydberg atoms have been
intensively investigated, mainly using a perturbative approach
\cite{Singer2005,Reinhard2007,Beguin2013}. The dispersion coefficients of the type
$C_6^{ss}$, $C_6^{pp}$ and $C_6^{dd}$, are calculated and
listed for homonuclear dimers of alkali metal atoms in the $ns-ns$, $np-np$
and $nd-nd$ states, where both atoms are in the same state. These coefficients are
of importance for experimental and theoretical applications in strongly
interacting Rydberg gases, especially for the current cold and ultracold Rydberg
atom experiments, e.g., in the implementation of dipole blockade phenomena for quantum
information processing \cite{Saffman2010,Low2012,Weimer2010}. But coefficients of the mixed
type in which atoms are in different electronic states, e.g., $C_6^{sp}$, are
not much emphasized, while they are of importance for processes that involve resonant energy
transfer \cite{Zoubi2014}.

In the present paper we study the influence of the formation of excitons on
van der Waals interactions among Rydberg atoms. We check the validity of using
the simple model of two-level systems to describe excitons in
interacting Rydberg atoms. We start by developing simple models using three
and four atomic levels that provide a qualitative understanding of the different
interactions among Rydberg atoms. In treating $ns-ns$ and $np-np$ Rydberg atoms we extract the
limit in which atom-atom interactions have the forms $C_6^{ss}/R^6$ and
$C_6^{pp}/R^6$. For the case of $ns-np$ Rydberg atoms we derive
the condition in which the interactions can be described by $C_3^{sp}/R^3$ and
$C_6^{sp}/R^6$ terms. Afterwards, we exploit perturbation theory in order to get quantitative
values for all $C_3$ and $C_6$ coefficients by summing over all the allowed atomic states. We treat Rydberg alkali metal atoms, and as an example we present the values for
lithium atoms.

We emphasize here van der Waals interactions of the mixed
type with coefficient $C_6^{sp}$, and examine the effect of resonant dipole-dipole transfer on van der Waals interactions. The point is that as the two states $|ns,np\rangle$ and
$|np,ns\rangle$ are degenerate, perturbation theory implies the removal of
these degeneracy. But the states are coupled by resonant dipole-dipole
interactions and the diagonalization mixes and splits them to yield symmetric and antisymmetric orthogonal
states. Then we use these orthogonal states as the zero
order eigenstates in the perturbative calculation of the van der Waals
interactions. We found that as the symmetric-antisymmetric splitting
energy is $R$-dependent new resonances appear as the interatomic distance decreases.

The paper is organized as follows. In section 2 we present a qualitative study
of van der Waals interactions among two Rydberg atoms using multi-level simple models. Quantitative derivations
of van der Waals interactions appear in
section 3 using perturbation theory. Section 4 contains
calculations of van der Waals coefficients for alkali Rydberg atoms, and the
results are presented for lithium atoms in section 5. A summary is given in section
6. The appendix includes the angular parts of the dipole moment matrix elements.

\section{Rydberg Atom Interactions: Simple Models}

We start by treating Rydberg atoms using models of two, three and
four-level systems. The derivations provide a qualitative
understanding of the type of interactions that can appear among Rydberg atoms
and pave the way towards the quantitative treatment presented in the next
section. We examine the validity of using two-level systems for describing
excitons in
Rydberg atoms. The model of two-level systems is widely used to describe Frenkel excitons in
organic solids involving lowest excited states where van der Waals interactions result in
small energy shifts \cite{Davydov1971,Agranovich2009}. But for Rydberg atoms van der Waals forces are significant
and the appearance of resonances is probable. Hence we examine the validity of
using the simple model of two-level systems for Rydberg atoms. Our objective in this section is to extend the simple model into
multi-level systems, which are necessary in order to exploit the appearance of resonances. We show
that far from resonances one can keep the simple model but in including energy
shifts due to van der Waals interactions, which implies quantitative calculations
of the van der Waals coefficients that we calculate in the next section using
perturbation theory.

\subsection{Two atoms in the same Rydberg state}

Let us assume the two atoms to be in the same internal atomic level, say in
the $(nl)$ state. The two-atom state is $|\pi\rangle=|nl,nl\rangle$ of energy
$E_i=2E_{nl}$. We assume a single channel to be close to resonance with
this state. Namely, we consider a process of the transfer type $nl+nl\leftrightarrow
n^{\prime}l^{\prime}+n^{\prime\prime}l^{\prime\prime}$. We have two possible degenerate final states
$|\rho_1\rangle=|n^{\prime}l^{\prime},n^{\prime\prime}l^{\prime\prime}\rangle$ and
$|\rho_2\rangle=|n^{\prime\prime}l^{\prime\prime},n^{\prime}l^{\prime}\rangle$, with the energy
$E_f=E_{n^{\prime}l^{\prime}}+E_{n^{\prime\prime}l^{\prime\prime}}$. The states have the energy detuning
$\Delta=E_f-E_i=E_{n^{\prime}l^{\prime}}+E_{n^{\prime\prime}l^{\prime\prime}}-2E_{nl}$. We
assume here coupling among the $|\pi\rangle$ state and each one of the
$|\rho_1\rangle$ and $|\rho_2\rangle$ states, with the coupling parameter $J$, which
we specify later. We neglect coupling among the $|\rho_1\rangle$ and
$|\rho_2\rangle$ states. The Hamiltonian that is restricted to the states defined
above can be written as
\begin{eqnarray}
H&=&E_i\ |\pi\rangle\langle\pi|+E_f\left(
  |\rho_1\rangle\langle\rho_1|+|\rho_2\rangle\langle\rho_2|\right) \nonumber \\
&+&J\left( |\pi\rangle\langle\rho_1|+|\pi\rangle\langle\rho_2|+|\rho_1\rangle\langle\pi|+|\rho_2\rangle\langle\pi|\right).
\end{eqnarray}
In matrix elements we have
\begin{eqnarray}
H=E_i\hat{\bf 1}+\left(\begin{tabular}{cccc}
$0$ & $J$ & $J$  \\
$J$  & $\Delta$ & $0$  \\
$J$  & $0$ & $\Delta$  \\
\end{tabular}
\right).
\end{eqnarray}
We diagonalize the matrix to get the characteristic equation
$(\lambda-\Delta)\left(\lambda^2-\lambda\Delta-2J^2\right)=0$, with the three solutions
\begin{equation}
\lambda_1=\frac{\Delta}{2}-\frac{1}{2}\sqrt{\Delta^2+8J^2},\
\lambda_2=\frac{\Delta}{2}+\frac{1}{2}\sqrt{\Delta^2+8J^2},\ \lambda_3=\Delta.
\end{equation}
In the far off-resonant limit, that is $\Delta\gg J$, we get
\begin{equation}
E_1\approx E_i-2\frac{J^2}{\Delta},\ E_2\approx E_f+2\frac{J^2}{\Delta},\ E_3=E_f.
\end{equation}
Transfer processes are induced by resonant dipole-dipole interaction that
have the form $J=-\frac{hC_3}{R^3}$, where $R$ is the interatomic distance, and $C_3$ will be
calculated in details later. We can also define
$D=-2\frac{(hC_3)^2}{\Delta}\frac{1}{R^6}$, which is of the van der Waals type. In the general case we have $D=-\frac{hC_6}{R^6}$, where one need to calculate the van der Waals coefficient
$C_6$ for each specific states by considering all possible atomic transitions, which is the main aim in the next sections. Next we consider the case of $(ns)$ state in details.

\begin{figure}[t]
\centerline{\epsfxsize=\columnwidth \epsfbox{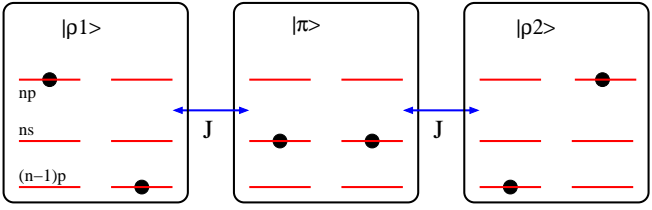}}
\caption{The three states $|\pi\rangle$, $|\rho_1\rangle$,
  and $|\rho_2\rangle$ are illustrated, with the coupling among
  them.}
\label{N2SS}
\end{figure}

In the case of $|\pi\rangle=|ns,ns\rangle$ of energy
$E_i=2E_{ns}$, we have, e.g., the channel $ns+ns\leftrightarrow
np+(n-1)p$, that is $|\rho_1\rangle=|np,(n-1)p\rangle$ and
$|\rho_2\rangle=|(n-1)p,np\rangle$, with the energy
$E_f=E_{np}+E_{(n-1)p}$, and with the detuning
$\Delta_{ss}=E_{np}+E_{(n-1)p}-2E_{ns}$, (see \fref{N2SS}). The energy of the two interacting atoms is
$E_{ss}=2E_{ns}+D_{ss}$, with the van der Waals interaction
$D_{ss}=-2\frac{\left(hC_3^{ss}\right)^2}{\Delta_{ss}}\frac{1}{R^6}$. In
the general case we can write $D_{ss}=-\frac{hC^{ss}_6}{R^6}$. Later on we calculate the
$C_3^{ss}$ and $C^{ss}_6$ coefficients in details. Similar consideration holds
for the case of two $(np)$-state atoms.

\subsection{Two atoms in different Rydberg states}

\begin{figure}[t]
\centerline{\epsfxsize=6cm \epsfbox{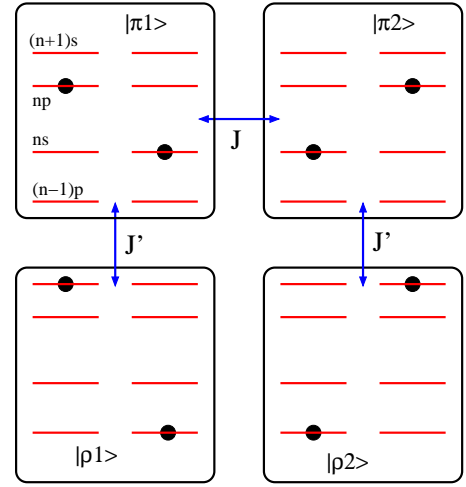}}
\caption{The four states $|\pi_1\rangle$, $|\pi_2\rangle$, $|\rho_1\rangle$,
  and $|\rho_2\rangle$ are illustrated, with the coupling among
  them.}
\label{N24}
\end{figure}

Here we consider the case in which the atoms are in different states, e.g., one atom in the $(ns)$-state and the other
in $(np)$-state. We get two degenerate states $|\pi_1\rangle=|np,ns\rangle$
and $|\pi_2\rangle=|ns,np\rangle$ of energy $E_i=E_{ns}+E_{np}$. We have,
e.g., the possible transfer channel of $ns+np\leftrightarrow
(n-1)p+(n+1)s$. Further, we consider the two degenerate states $|\rho_1\rangle=|(n+1)s,(n-1)p\rangle$ and
$|\rho_2\rangle=|(n-1)p,(n+1)s\rangle$, of energy
$E_f=E_{(n+1)s}+E_{(n-1)p}$, (see \fref{N24}). Their energy detuning compared
to $|\pi\rangle$ state is
$\Delta_{sp}=E_{(n+1)s}-E_{ns}+E_{(n-1)p}-E_{np}$. The states $|\pi_1\rangle=|np,ns\rangle$
and $|\pi_2\rangle=|ns,np\rangle$ are coupled by the resonant dipole-dipole
interaction of strength $J$, where $J=-\frac{hC_3^{sp}}{R^3}$. The state
$|\pi_1\rangle=|np,ns\rangle$ is coupled to $|\rho_1\rangle=|(n+1)s,(n-1)p\rangle$ by the resonant dipole-dipole
interaction of strength $J^{\prime}$, and the state $|\pi_2\rangle=|ns,np\rangle$ is coupled to $|\rho_2\rangle=|(n-1)p,(n+1)s\rangle$ with the same parameter $J^{\prime}$. We neglect the resonant dipole-dipole
interaction among the $|\rho_1\rangle=|(n+1)s,(n-1)p\rangle$ and
$|\rho_2\rangle=|(n-1)p,(n+1)s\rangle$ states, due to their small amplitudes.

The Hamiltonian can be then written as
\begin{eqnarray}
H&=&E_i\left(
  |\pi_1\rangle\langle\pi_1|+|\pi_2\rangle\langle\pi_2|\right)+E_f\left(
  |\rho_1\rangle\langle\rho_1|+|\rho_2\rangle\langle\rho_2|\right) \nonumber \\
&+&J\left( |\pi_1\rangle\langle\pi_2|+|\pi_2\rangle\langle\pi_1|\right) \nonumber \\
&+&J^{\prime}\left( |\pi_1\rangle\langle\rho_1|+|\pi_2\rangle\langle\rho_2|+|\rho_1\rangle\langle\pi_1|+|\rho_2\rangle\langle\pi_2|\right).
\end{eqnarray}
In matrix elements we get
\begin{eqnarray}
H=E_i\hat{\bf 1}+\left(\begin{tabular}{cccc}
$0$ & $J$   & $J'$ & $0$ \\
$J$   & $0$ & $0$ & $J'$  \\
$J'$   & $0$  & $\Delta_{sp}$ & $0$  \\
$0$   & $J'$  & $0$ & $\Delta_{sp}$  \\
\end{tabular}
\right).
\end{eqnarray}
We diagonalize the Hamiltonian, to get the characteristic equations
\begin{equation}
\left(\lambda-\Delta_{sp}\right)^2\left(\lambda^2-J^2\right)-2\lambda\left(\lambda-\Delta_{sp}\right)J^{\prime\ 2}+J^{\prime\ 4}=0,
\end{equation}
with the four solutions
\begin{eqnarray}
\lambda_{1,2}&=&\frac{\Delta_{sp}\mp
  J}{2}-\frac{1}{2}\sqrt{\left(\Delta_{sp}\pm J\right)^2+4J^{\prime\ 2}}, \nonumber \\
\lambda_{3,4}&=&\frac{\Delta_{sp}\mp
  J}{2}+\frac{1}{2}\sqrt{\left(\Delta_{sp}\pm J\right)^2+4J^{\prime\ 2}}.
\end{eqnarray}
In the far off-resonant limit, that is $\Delta_{sp}\gg J'$, we get the energies
\begin{eqnarray}
E_{a,b}&=&E_{ns}+E_{np}\mp J-\frac{J^{\prime\ 2}}{\Delta_{sp}\pm J}, \nonumber \\
E_{c,d}&=&E_{(n+1)s}+E_{(n-1)p}+\frac{J^{\prime\ 2}}{\Delta_{sp}\pm J}.
\end{eqnarray}
As far as $\Delta_{sp}\gg J$ the shifts are of the van der Waals type, where $D_{sp}=-J^{\prime\ 2}/\Delta_{sp}$. In general we can write
$D_{sp}=-\frac{hC_6^{sp}}{R^6}$, where the $C_6$ coefficient will be calculated
later.

\subsection{Two-level atoms}

\begin{figure}[t]
\centerline{\epsfxsize=7cm \epsfbox{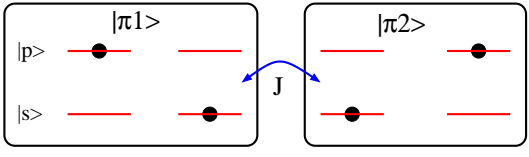}}
\caption{The two states $|\pi_1\rangle$ and $|\pi_2\rangle$ are illustrated
  with the coupling $J$ among them.}
\label{N2}
\end{figure}

In the light of this result, we can go one step back and start with two
effective two-level atoms including van der Waals interactions. We
again treat the case of one atom in the
$(ns)$ Rydberg state, and the other in the $(np)$ state, as seen in \fref{N2}. The atoms are
separated by the distance $R$. We have two states
$|\pi_1\rangle=|np,ns\rangle$ and
$|\pi_2\rangle=|ns,np\rangle$, and they are degenerate with
the energy $E_0=E_{ns}+E_{np}+D_{sp}$, where we include the van der Waals
interaction among the two atoms $D_{sp}=-hC_6^{sp}/R^6$. The energy transfer parameter among the two states
is as before $J=-hC_3^{sp}/R^3$. These consideration exactly fits with the above
four-level model in the limit of $\Delta_{sp}\gg J$, where the detuning
$\Delta_{sp}$ is much larger than the symmetric-antisymmetric splitting.

The Hamiltonian is now written as
\begin{equation}
H=E_0\left(|\pi_{1}\rangle\langle\pi_{1}|+|\pi_{2}\rangle\langle\pi_{2}|\right)+J\left( |\pi_{1}\rangle\langle\pi_{2}|+|\pi_{2}\rangle\langle\pi_{1}|\right).
\end{equation}
The Hamiltonian can be diagonalized by using the collective states
\begin{equation}
|\psi_a\rangle=\frac{|\pi_1\rangle+|\pi_2\rangle}{\sqrt{2}},\ |\psi_b\rangle=\frac{|\pi_1\rangle-|\pi_2\rangle}{\sqrt{2}},
\end{equation}
which gives
\begin{equation}
H=E_a\ |\psi_a\rangle\langle\psi_a|+E_b\ |\psi_b\rangle\langle\psi_b|,
\end{equation}
with the energies $E_a=E_0+J$ and $E_b=E_0-J$, which fit exactly with the above derived energies.

As other states can be close to resonance with
the considered states, the picture of two-level atoms breaks down at short
interatomic distances of several $\mu m$. Then, the excitonic picture implies
multi-level systems, as beside the considered states, e.g. $ns$ and $np$, the
formalism must include all close to resonance states, e.g $(n+1)s$ and
$(n-1)p$, as treated before. The model of two-level systems can be reserved for larger
interatomic distances when other states are off-resonance, but in including van der Waals interactions among the atoms. We adopted this
direction in treating $N$ Rydberg atoms in our previous paper
\cite{Zoubi2014}. At much larger interatomic distances of tenth of $\mu m$ the van
der Waals interactions are negligible.

After this qualitative study of the resonant dipole-dipole and van der Waals
interactions, we move to quantitative calculations of the different $C_3$ and
$C_6$ coefficients.

\section{Dimeric Energies in Perturbation Theory}

We give first a general presentation of a perturbative treatment for
the electrostatic interactions among Rydberg atoms. For a system of two atoms
the Hamiltonian reads $\hat{H}=\hat{H}_{1}+\hat{H}_{2}+\hat{V}$, where $\hat{H}_{\alpha}$ is the $\alpha$ independent atom Hamiltonian, and $\hat{V}$ is the
interaction among the two atoms. We assume the independent atom eigenstates to be known
and given by
$\hat{H}_{\alpha}|\phi_{\alpha}^i\rangle=E_{\alpha}^i|\phi_{\alpha}^i\rangle$,
where the eigenstates are orthonormal with
$\langle\phi_{\beta}^j|\phi_{\alpha}^i\rangle=\delta_{\alpha\beta}\delta_{ij}$,
and $\sum_i|\phi_{\alpha}^i\rangle\langle\phi_{\alpha}^i|=\hat{\bf 1}$.

The lowest order interaction among neutral atoms is the dipole-dipole one, where
\begin{equation}
\hat{V}=\frac{1}{4\pi\epsilon_0R^3}\left\{\left(\hat{\mbox{\boldmath$\mu$}}_{1}\cdot\hat{\mbox{\boldmath$\mu$}}_{2}\right)-3\frac{\left(\hat{\mbox{\boldmath$\mu$}}_{1}\cdot{\bf R}\right)\left(\hat{\mbox{\boldmath$\mu$}}_{2}\cdot{\bf R}\right)}{R^5}\right\},
\end{equation}
with ${\bf R}={\bf R}_{1}-{\bf R}_{2}$, and the interatomic distance is $R=|{\bf R}|$. Here
$\hat{\mbox{\boldmath$\mu$}}_{\alpha}$ is the dipole operator at atom $\alpha$. We assume
no permanent dipoles to exist, and hence
$\mbox{\boldmath$\mu$}_{\alpha}^{ii}=\langle\phi_{\alpha}^i|\hat{\mbox{\boldmath$\mu$}}_{\alpha}|\phi_{\alpha}^i\rangle=0$. We have
$\mbox{\boldmath$\mu$}_{\alpha}^{ij}=\langle\phi_{\alpha}^i|\hat{\mbox{\boldmath$\mu$}}_{\alpha}|\phi_{\alpha}^j\rangle$
with $(i\neq j)$. The atoms are taken to be localized along the $z$-axis, which
is also the quantization axis, hence
\begin{equation}
\hat{V}=\frac{1}{4\pi\epsilon_0R^3}\left\{\hat{\mu}_{1x}\hat{\mu}_{2x}+\hat{\mu}_{1y}\hat{\mu}_{2y}-2\hat{\mu}_{1z}\hat{\mu}_{2z}\right\}.
\end{equation}
We apply perturbation
theory in order to calculate the resonant dipole-dipole and van der Waals
coefficients by deriving the corrections to each dimer energy.

We write the perturbative terms up to the second order explicitly where the two atoms are assumed
to be in the states $i$ and $j$. The lowest order correction to the free
energy is
$J^{ij,kl}=\langle\phi_{1}^i,\phi_{2}^j|\hat{V}|\phi_{1}^k,\phi_{2}^l\rangle$,
where for the dimer we use a product basis
$|\phi_{1}^i,\phi_{2}^j\rangle=|\phi_{1}^i\rangle|\phi_{2}^j\rangle$. This
term give rise to resonant dipole-dipole interaction that is responsible
for the energy transfer among atoms at different internal states. When the two atoms are in
the same $i$-th quantum state one has $\langle\phi_{1}^i,\phi_{2}^i|\hat{V}|\phi_{1}^i,\phi_{2}^i\rangle=0$. Only
terms of $(i\neq k)$ and $(j\neq l)$ are nonzero, with the appropriate
selection rules. We define $J^{ij,kl}=-\frac{hC_3^{ij,kl}}{R^3}$, where now
\begin{equation}
C_3^{ij,kl}=-\frac{1}{4\pi\epsilon_0}\left\{\mu_{1x}^{ik}\mu_{2x}^{jl}+\mu_{1y}^{ik}\mu_{2y}^{jl}-2\mu_{1z}^{ik}\mu_{2z}^{jl}\right\},
\end{equation}
and the transition dipole matrix element is defined by $\mu_{\alpha
  r}^{ij}=\langle\phi_{\alpha}^i|\hat{\mu}_{\alpha r}|\phi_{\alpha}^j\rangle$,
with $(r=x,y,z)$.

Resonances can appear among the dimer states, e.g., for
$|\phi_{1}^i,\phi_{2}^j\rangle$ and $|\phi_{1}^k,\phi_{2}^l\rangle$ states
with the energy $E_{1}^k+E_{2}^l=E_{1}^i+E_{2}^j$, which implies the
use of degenerate perturbation theory. In the following we present separately
the degenerate and non-degenerate perturbation theory.

\subsection{Non-degenerate states}

For non degenerate states
$|\phi_{1}^i,\phi_{2}^j\rangle$  the first order term
$\langle\phi_{1}^i,\phi_{2}^j|\hat{V}|\phi_{1}^i,\phi_{2}^j\rangle$
vanishes, because the atoms do not have a permanent dipole.
The second order term is
\begin{equation}
D^{ij}=-\sum_{k,l}\frac{\langle\phi_{1}^i,\phi_{2}^j|\hat{V}|\phi_{1}^k,\phi_{2}^l\rangle\langle\phi_{1}^k,\phi_{2}^l|\hat{V}|\phi_{1}^i,\phi_{2}^j\rangle}{E_{1}^k-E_{1}^i+E_{2}^l-E_{2}^j},
\end{equation}
which results in van der Waals interactions. 
Since $\hat{V}$ has a $1/R^3$ distance dependence, we can write $D^{ij}=-\frac{hC_6^{ij}}{R^6}$, where we have defined
\begin{equation}
C_6^{ij}=\left(\frac{1}{4\pi\epsilon_0}\right)^2\sum_{k,l}\frac{\left|\mu_{1x}^{ik}\mu_{2x}^{jl}+\mu_{1y}^{ik}\mu_{2y}^{jl}-2\mu_{1z}^{ik}\mu_{2z}^{jl}\right|^2}{E_{1}^k-E_{1}^i+E_{2}^l-E_{2}^j}.
\end{equation}
The present perturbation theory can break down due to the appearance of
resonances at $E_{1}^k+E_{2}^l=E_{2}^j+E_{1}^i$, and one appeals to other
methods, e.g., to the direct
Hamiltonian diagonalization \cite{Schwettmann2006}.

\subsection{Degenerate States}

We now consider states $|\phi_{1}^i,\phi_{2}^j\rangle$ for which a state $|\phi_{1}^k,\phi_{2}^l\rangle$ with the same energy exists, i.e.\ $E_i+E_j = E_k+E_l$, and for which also $J^{ij,kl}=\langle\phi_{1}^i,\phi_{2}^j|\hat{V}|\phi_{1}^k,\phi_{2}^l\rangle \ne 0$.
Then we have to use degenerate perturbation theory and first diagonalize the
degenerate subspace. Here we concentrate in degenerate states of the type
$|\phi_{1}^i,\phi_{2}^j\rangle$ and $|\phi_{1}^j,\phi_{2}^i\rangle$. The 'correct' zero order states have the form
\begin{equation}
|\phi_\pm^{ij}\rangle =|\phi_{1}^i,\phi_{2}^j\rangle \pm  |\phi_{1}^j,\phi_{2}^i\rangle,
\end{equation}
with energy
\begin{equation}
E_{\pm}^{ij}=E_1^i+E_2^j \pm J^{ij}.
\end{equation}
The first order contributions then vanish and the second order term, which has a $1/R^6$ behavior is calculated by
\begin{equation}
D^{ij}_{\pm}=-\sum_{k,l}\frac{|\langle\phi_{\pm}^{ij}|\hat{V}|\phi^{k}_1,\phi^l_2\rangle|^2}{E^{k}_1+E^{l}_2-E_{\pm}^{ij}},
\end{equation}
where the sum over $k$ and $l$ includes all states but
$|\phi_{\pm}^{ij}\rangle$. Note that resonances appear at
$E^{k}_1+E^{l}_2=E_{\pm}^{ij}$ and then a special treatment is required.

\section{Interactions between Alkali Rydberg Atoms}

We calculate the $C_3$ and $C_6$ coefficients for specific
cases that include the $|\phi^{ns}\rangle$ and $|\phi^{np}\rangle$ states, and we concentrate in interactions among alkali Rydberg atoms. We
use the nondegenerate perturbation theory in order to calculate $C_6^{ss}$ and
$C_6^{pp}$ coefficients. Next we calculate $C_6^{sp}$
coefficients, and as here a resonance appears between the states
$|\phi_{1}^{ns},\phi_{2}^{np}\rangle$ and
$|\phi_{1}^{np},\phi_{2}^{ns}\rangle$ we use degenerate perturbation
theory. Let us first present the atomic states for alkali Rydberg atoms.

\subsection{Atomic eigenstates}

The atomic state is given by the wave function $\psi_{nlm}(r,\theta,\phi)=R_{nl}(r)Y_{lm}(\theta,\phi)$, where $R_{nl}(r)$
is the radial wavefunction and $Y_{lm}(\theta,\phi)$ is the spherical harmonic function. In general the
dipole moment matrix element can be factorized as
$\mu_{r}^{nlm,n'l'm'}=e\tilde{\mu}_{r}^{lm,l'm'}{\cal I}^{nl}_{n'l'}$, where the radial integral is
\begin{equation}
{\cal I}^{nl}_{n'l'}=\int_0^{\infty}dr\ r^3R_{n'l'}R_{nl},
\end{equation}
and the angular part is
\begin{equation}
\tilde{\mu}_{i}^{lm,l'm'}=\int_0^{\pi}d\theta\int_0^{2\pi}d\phi \sin\theta
\ Y^{\ast}_{lm}(\theta,\phi)\hat{\bf e}_iY_{l'm'}(\theta,\phi).
\end{equation}
Here $\hat{\bf e}_i$ are the axis unit vectors with
\begin{equation}
\hat{\bf e}_x=\sin\theta\cos\phi,\ \hat{\bf e}_y=\sin\theta\sin\phi,\ \hat{\bf e}_z=\cos\theta.
\end{equation}

The radial wave function for alkali atoms in atomic units is given by
\begin{eqnarray}
R_{nl}(r)&=&\frac{2}{n^{\star
    2}}\sqrt{\frac{(n^{\star}-l^{\star}-1)!}{(n^{\star}+l^{\star})!}}
\nonumber \\
&\times&\left(\frac{2r}{n^{\star}}\right)^{l^{\star}}e^{-r/n^{\star}}L^{2l^{\star}+1}_{n^{\star}-l^{\star}-1}\left(\frac{2r}{n^{\star}}\right),
\end{eqnarray}
where $n^{\star}=n-\delta_{nl}$ and $l^{\star}=l-\delta_{nl}+I(l)$, with the
quantum defect $\delta_{nl}=\delta_0^l+\frac{\delta_2^l}{(n-\delta_0^l)^2}$, and
$I(l)$ is the nearest integer below or equal to
$\delta_{nl}$. For the energy we have $E_{nl}=\frac{-1}{2(n-\delta_{nl})^2}$. Moreover we neglect fine and hyperfine splitting, which is a good
approximation for lithium atoms that we present later.

\subsection{$C_6^{ss}$ and $C_6^{pp}$ Coefficients}

We present here detailed calculations of $C_6^{ss}$ and $C_6^{pp}$
coefficients for atoms
in the $(ns)$ and  $(np)$ states. We concentrate on alkali metal
atoms excited to Rydberg states and hence  we deal with a single electronic
state at each atom that is represented by three quantum numbers $(n,l,m)$, where
$n=1,2,\cdots$, $l=0,\cdots,n-1$, and $m=-l,\cdots,+l$.

We write
the expression for the van de Waals coefficients as follows
\begin{eqnarray}
C_{6}^{ns,ns}&=&\left(\frac{e^2}{4\pi\epsilon_0}\right)^2A_{ns}^{ns},
\nonumber \\
C_{6}^{npm,np\bar{m}}&=&\left(\frac{e^2}{4\pi\epsilon_0}\right)^2A_{np\bar{m}}^{npm},
\end{eqnarray}
where we defined
\begin{equation}
A_{n\bar{l}\bar{m}}^{nlm}=\sum_{n'l',n''l''}^{\prime}\left|\wp_{\bar{l}\bar{m},l''}^{lm,l'}\right|^2\frac{|{\cal
    I}^{nl}_{n'l'}{\cal
    I}^{n\bar{l}}_{n''l''}|^2}{\Delta^{nl,n\bar{l}}_{n'l',n''l''}},
\end{equation}
and the prime over the summation indicates that $nl\neq n'l'$ and
$n\bar{l}\neq n''l''$. We have
\begin{equation}
\Delta^{nl,n\bar{l}}_{n'l',n''l''}=E_{n'l'}+E_{n''l''}-E_{nl}-E_{n\bar{l}},
\end{equation}
and
\begin{eqnarray}\label{AngSSPP}
|\wp_{\bar{l}\bar{m},l''}^{lm,l'}|^2&=&\sum_{m',m''}\left|\tilde{\mu}_{x}^{lm,l'm'}\tilde{\mu}_{x}^{\bar{l}\bar{m},l''m''}+\tilde{\mu}_{y}^{lm,l'm'}\tilde{\mu}_{y}^{\bar{l}\bar{m},l''m''}\right. \nonumber \\
&-&\left.2\tilde{\mu}_{z}^{lm,l'm'}\tilde{\mu}_{z}^{\bar{l}\bar{m},l''m''}\right|^2.
\end{eqnarray}
As we treat two identical atoms, we dropped the atom index. We calculate the radial parts, ${\cal I}$, of the $A$ parameters numerically, and the
angular parts, $\left|\wp\right|^2$, are calculated analytically and presented
in the appendix.

We aim now to calculate the $C_3$ coefficient, which can be written as
\begin{equation}
C_3^{s,pm}=\frac{-1}{4\pi\epsilon_0}\wp^{s,pm}\left|{\cal I}^{np}_{ns}\right|^2,
\end{equation}
where
\begin{equation}
\wp^{s,pm}=\tilde{\mu}_{x}^{s,pm}\tilde{\mu}_{x}^{pm,s}+\tilde{\mu}_{y}^{s,pm}\tilde{\mu}_{y}^{pm,s}-2\tilde{\mu}_{z}^{s,pm}\tilde{\mu}_{z}^{pm,s},
\end{equation}
and here we have $\wp^{s,p0}=-\frac{2}{3}$ and $\wp^{s,p\pm1}=\frac{1}{3}$.

\subsection{$C_6^{sp}$ Coefficients}

One needs to be careful when calculating the $C_6^{sp}$ coefficients, as the states
$|ns,np\rangle$ and $|np,ns\rangle$ are degenerate, and perturbation theory
implies the removal of this degeneracy, as we presented before. The two states are coupled by the
resonant dipole-dipole interaction, which yields two diagonal states
of symmetric and antisymmetric mixing, where these orthogonal states used as the zero order basis for the second order
perturbation theory.

The diagonal eigenstates are
\begin{equation}
|\pm\rangle=\frac{|ns,npm\rangle\pm|npm,ns\rangle}{\sqrt{2}},
\end{equation}
with the diagonal energies
\begin{equation}
E_{n}^{\pm}(R)=E_{ns}+E_{np}\pm J^{ns,npm}(R),
\end{equation}
where
\begin{equation}
J^{ns,npm}(R)=-\frac{hC_3^{ns,npm}}{R^3}.
\end{equation}
The diagonal states have splitting energy of $2J^{ns,npm}(R)$. We aim to calculate
\begin{eqnarray}
D_{\alpha}^{ns,npm}&=&-\left(\frac{1}{4\pi\epsilon_0R^3}\right)^2 \nonumber \\
&\times&\sum_{k,l}\frac{\left|\mu_x^{\alpha k}\mu_x^{\alpha l}+\mu_y^{\alpha k}\mu_y^{\alpha l}-2\mu_z^{\alpha k}\mu_z^{\alpha l}\right|^2}{E_k+E_l-E^{\alpha}_n(R)},
\end{eqnarray}
with $(\alpha=\pm)$, $(k=n'l'm')$ and $(l=n''l''m'')$.

The calculation yields
\begin{eqnarray}
&&D_{\pm}^{ns,npm}=-2\left(\frac{e^2}{4\pi\epsilon_0R^3}\right)^2 \nonumber \\
&\times&\sum_{n'l',n''l''}^{\prime}\left|\wp_{pm,l''}^{s,l'}\right|^2\frac{|{\cal
    I}^{ns}_{n'l'}{\cal I}^{np}_{n''l''}|^2}{\Delta^{ns,np}_{n'l',n''l''}\mp J_n^{s,pm}(R)}.
\end{eqnarray}

The $R$ dependence in the denominator through $J_n^{s,pm}(R)$ can bring new
resonances as atoms approach each other and then the present perturbation
theory breaks
down. Event hough, far from resonance, that is
$\Delta^{ns,np}_{n'l',n''l''}\gg J_n^{s,pm}(R)$, the $C_6^{sp}$ coefficients can be safely
defined by $D^{sp}=-hC_6^{sp}/R^6$, where
\begin{equation}
C_{6}^{ns,np\bar{m}}=\left(\frac{e^2}{4\pi\epsilon_0}\right)^2A_{np\bar{m}}^{ns}.
\end{equation}
As before the radial parts are calculated numerically, and the angular parts
calculated analytically and presented in the appendix.

\section{Values for Lithium Atom}

We calculate now all $C_6$ and $C_3$ coefficients for lithium atoms, by using the
known defect parameters \cite{Lorentzen1983}. For $ns$ we have
$\delta_0^s=0.399$ and $\delta_2^s=0.03$, for $np$ we have
$\delta_0^p=0.0473$ and $\delta_2^p=-0.026$, and for $nd$ we have $\delta_0^d=0.002$ and $\delta_2^d=-0.015$.

Converting the above $A$ parameters to SI units requires the change $A\rightarrow
\left(\frac{4\pi\epsilon_0a_0^5}{e^2}\right)A$, as we have a factor of
$\frac{e^2}{4\pi\epsilon_0a_0}$ from the energy, and a factor of $a_0$ from
the radial function ${\cal I}$. We obtain $C_{6}^{ns,ns}=BA_{ns}^{ns}$ and
$D_{6}^{np\bar{m},npm}=BA_{np\bar{m}}^{npm}$, where $B=\frac{e^2a_0^5}{4\pi\epsilon_0}$. Using $\epsilon_0\approx 0.0055\ e/(V\AA)$, and $a_0\approx 0.53\ \AA$, we
get $B\approx 0.6\ eV\AA^6$, or $B=h\bar{B}$, with $h\approx 4.135\times
10^{-15}\ eVS$, we get $\bar{B}\approx 1.5\times 10^{-10}\ Hz(\mu m)^6$. Moreover, we have $C_3^{s,pm}=-F\wp^{s,pm}\left|{\cal I}^{np}_{ns}\right|^2$, where $F=\frac{a_0^2e^2}{4\pi\epsilon_0}$. We use $F=h\bar{F}$ with $\bar{F}\approx 10^{3}\ Hz(\mu m)^3$

\begin{figure}[t]
\centerline{\epsfxsize=6cm \epsfbox{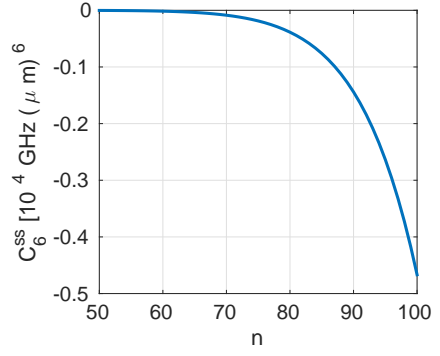}}
\caption{$C_{6}^{ss}$ as a
  function of $n$ for lithium atoms.}
\label{C6ssLiHC}
\end{figure}

\begin{figure}[t]
\centerline{\epsfxsize=6cm \epsfbox{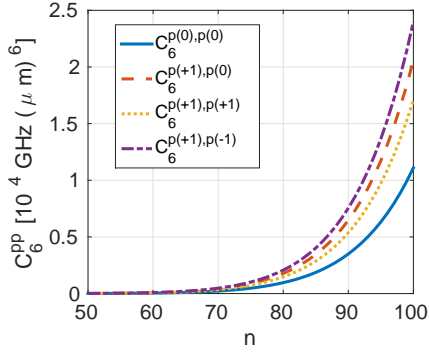}}
\caption{$C_{6}^{p0,p0}$, $C_{6}^{p0,p\pm 1}$ and $C_{6}^{p\pm 1,p\pm 1}$ as a
  function of $n$ for lithium atoms.}
\label{C6ppLi}
\end{figure}

\begin{figure}[t]
\centerline{\epsfxsize=6cm \epsfbox{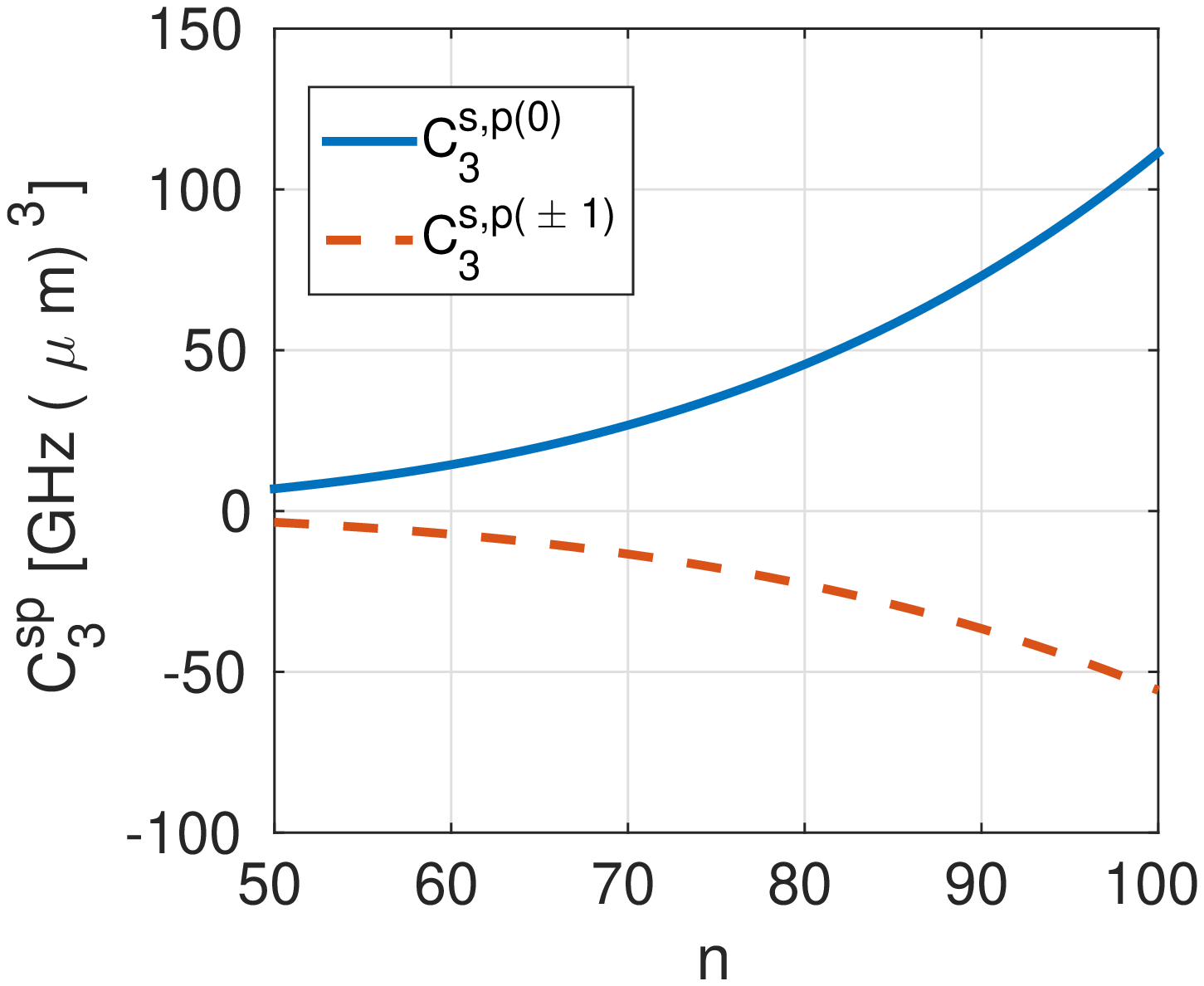}}
\caption{$C_{3}^{s,p0}$, and $C_{6}^{s,p\pm1}$ as a
  function of $n$ for lithium atoms.}
\label{C3spLi}
\end{figure}

We plot now the results for all $C_6^{ss}$, $C_6^{pp}$ and $C_3^{sp}$ coefficients ranging from
$n=50$ up to $n=100$. In \fref{C6ssLiHC} we plot $C_{6}^{ss}$, and in \fref{C6ppLi} we plot
$C_{6}^{p0,p0}$, $C_{6}^{p0,p\pm 1}$ and $C_{6}^{p\pm 1,p\pm 1}$. Here
$C_6^{ss}$ coefficients are negative that lead to repulsive van der Waals forces, and
$C_6^{pp}$ are positive that lead to attractive ones. The results for
$C_3^{s,p0}$ and $C_3^{s,p\pm1}$ are plotted in \fref{C3spLi}. $C_3^{s,p0}$
coefficients are positive that lead to attractive resonant dipole-dipole
interactions, and $C_3^{s,p\pm1}$ are negative that lead to repulsive ones.

\begin{figure}[t]
\centerline{\epsfxsize=6cm \epsfbox{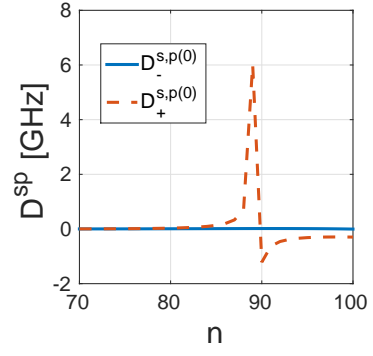}}
\caption{$D_{+}^{s,p0}$ and $D_{-}^{s,p0}$ as a
  function of $n$ for $R=7 \mu m$ of lithium atoms.}
\label{D0SP7}
\end{figure}

\begin{figure}[t]
\centerline{\epsfxsize=6cm \epsfbox{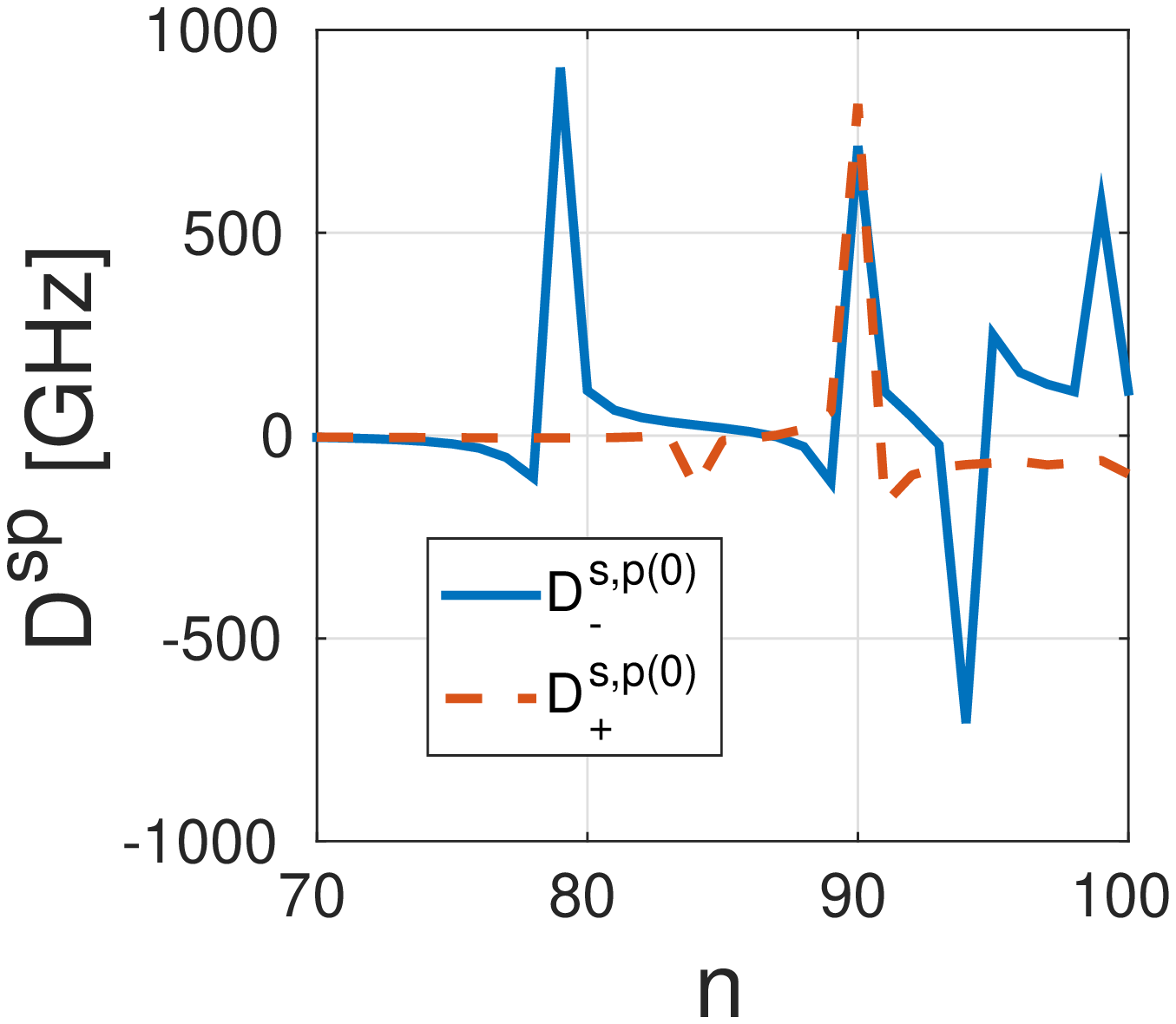}}
\caption{$D_{+}^{s,p0}$ and $D_{-}^{s,p0}$ as a
  function of $n$ for $R=2 \mu m$ of lithium atoms.}
\label{D0SP2}
\end{figure}

\begin{figure}[t]
\centerline{\epsfxsize=6cm \epsfbox{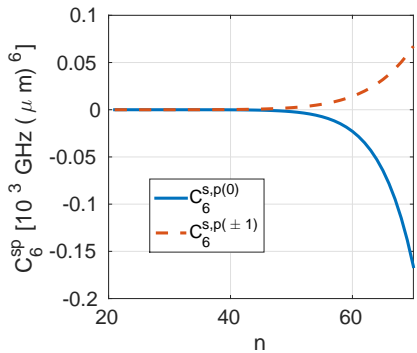}}
\caption{$C_6^{s,p(0)}$ and $C_6^{s,p(\pm 1)}$ as a
  function of $n$ for lithium atoms.}
\label{C6SP}
\end{figure}

Now we present the results for $D_{\pm}^{sp}$ energies that derived in using the diagonal states as zero
order in perturbation theory for lithium atoms. The main feature of
$D_{\pm}^{sp}$ is in their $R$
dependent. We present the plots for different interatomic distances. In \fref{D0SP7} we plot $D_{+}^{s,p0}$ and
$D_{-}^{s,p0}$ for $R=7 \mu m$, and in \fref{D0SP2} for $R=2 \mu
m$. It is clear that resonances start to appear at smaller $n$ as atoms
approach each other. The $D_{+}^{s,p}$ and $D_{-}^{s,p}$ energies are
different due to the symmetric-antisymmetric energy splitting, but for small
$n$ as the splitting is much smaller than the energy detuning then the
coefficients become equal, and in this limit one can define $C_6^{sp}$
coefficients. In \fref{C6SP} we plot $C_6^{s,p(0)}$ and $C_6^{s,p(\pm 1)}$
coefficients from $(n=20)$ up to $(n=70)$. In this region no resonances appear down to
small interatomic distances of $2\ \mu m$, as it is clear from the above
plots. Here the $C_6^{s,p(0)}$ coefficients are negative for $n>30$ and then the van der
Waals interaction is repulsive, but the $C_6^{s,p(\pm 1)}$ coefficients are
positive and the interaction is attractive.

\section{Conclusions}

In summary we investigated van der Waals interactions among two Rydberg atoms
for the case where the atoms are in different internal states. For example, we treated
in details the case with one atom in the $(ns)$-state and the other in
$(np)$-state. The simplest model for achieving a qualitative understanding of
the interactions is of four-level atoms, e.g., with $|ns\rangle$, $|np\rangle$,
$|n^{\prime}p\rangle$ and $|n^{\prime\prime}s\rangle$ states of energies $E_{ns}$, $E_{np}$, $E_{n^{\prime}p}$ and $E_{n^{\prime\prime}s}$, respectively. The two states $|ns,np\rangle$ and $|np,ns\rangle$ are
degenerate and couple by resonant dipole-dipole interaction of the type
$J_{sp}=-hC_3^{sp}/R^3$, which mixes and splits the states to give symmetric and
antisymmetric orthogonal ones. We show that van der Waals interaction among
the two atoms can be formulated as $D_{sp}=-hC_6^{sp}/R^6$ in the limit of
off-resonance, that is in the limit $|E_{ns}+E_{np}-E_{n^{\prime}p}-E_{n^{\prime\prime}s}|\gg|J_{sp}|$. In
this limit the model of two-level atoms is valid, but in including van der
Waals interactions.

Next we presented a quantitative calculation of van der Waals coefficients
among two alkali Rydberg atoms in considering all allowed electronic
states. Due to the resonance among the states $|ns,np\rangle$ and $|np,ns\rangle$ we
applied degenerate perturbation theory. The zero order states used in
perturbation theory are the orthogonal states
$|nsp\rangle_{\pm}=\frac{|ns,np\rangle\pm|np,ns\rangle}{\sqrt{2}}$ with
energies $E_{\pm}^{nsp}=E_{ns}+E_{np}\pm J_{nsp}(R)$. We found that due to the
symmetric-antisymmetric energy splitting dependence on $R$ new resonances
appear when atoms approach each other, and only far from these resonance one
can define $C_6^{sp}$ coefficients. Moreover, using non-degenerate perturbation theory we calculated all van der Waals
coefficients among alkali Rydberg atoms in the same state, mainly for the
$|ns,ns\rangle$ and $|np,np\rangle$ states, and we presented the results for
lithium atoms.

\section*{Acknowledgment}

The author thanks Alex Eisfeld and
Sebastian W{\"u}ster for very fruitful discussions. The author acknowledges
Max-Planck Institute for the Physics of Complex Systems in Dresden-Germany for
the hospitality through the visitor program grant, while big part of this work was
done.

\appendix

\section{Angular Part of Dipole Moment Matrix Elements}

In this appendix we present the angular part calculations of the dipole moment
matrix elements. Direct calculations yield the non-zero matrix elements
\begin{eqnarray}
\tilde{\mu}_{x}^{00,1\pm 1}=\frac{\mp 1}{\sqrt{6}}&,&\tilde{\mu}_{x}^{10,2\pm
  1}=\frac{\mp 1}{\sqrt{10}}, \nonumber \\
\tilde{\mu}_{x}^{1\pm 1,20}=\frac{\pm 1}{\sqrt{30}}&,&\tilde{\mu}_{x}^{1\pm 1,2\mp 2}=\frac{\mp 1}{\sqrt{5}} \nonumber \\
\tilde{\mu}_{y}^{00,1\pm 1}=\frac{-i}{\sqrt{6}}&,&\tilde{\mu}_{y}^{10,2\pm
  1}=\frac{-i}{\sqrt{10}}, \nonumber \\
\tilde{\mu}_{y}^{1\pm 1,20}=\frac{i}{\sqrt{30}}&,&\tilde{\mu}_{y}^{1\pm 1,2\mp
  2}=\frac{i}{\sqrt{5}}, \nonumber \\
\tilde{\mu}_{z}^{00,10}=\frac{1}{\sqrt{3}}&,&\tilde{\mu}_{z}^{10,20}=\sqrt{\frac{2}{15}},
\nonumber \\
\tilde{\mu}_{z}^{1\pm 1,2\mp 1}=\frac{-1}{\sqrt{5}},&&
\end{eqnarray}
and all the others vanish. The angular factors of equation (\ref{AngSSPP}) are
\begin{eqnarray}
&&\left|\wp_{s,p}^{s,p}\right|^2=\frac{2}{3}\ ,\ \left|\wp_{p0,s}^{s,p}\right|^2=\frac{4}{9}\
,\ \left|\wp_{p\pm 1,s}^{s,p}\right|^2=\frac{1}{9},
\nonumber \\
&&\left|\wp_{p0,d}^{s,p}\right|^2=\frac{14}{45}\ ,\ \left|\wp_{p\pm 1,d}^{s,p}\right|^2=\frac{19}{45},
\nonumber \\
&&\left|\wp_{p0,s}^{p0,s}\right|^2=\frac{4}{9}\ ,\ \left|\wp_{p-1,s}^{p1,s}\right|^2=\left|\wp_{p1,s}^{p-1,s}\right|^2=\frac{1}{9},
\nonumber \\
&&\left|\wp_{p0,s}^{p\pm 1,s}\right|^2=\left|\wp^{p0,s}_{p\pm 1,s}\right|^2=0\ ,\ 
\left|\wp_{p1,s}^{p1,s}\right|^2=\left|\wp_{p-1,s}^{p-1,s}\right|^2=0,
\nonumber \\
&&\left|\wp_{p0,d}^{p0,s}\right|^2=\frac{8}{45}\ ,\ \left|\wp_{p0,d}^{p\pm1,s}\right|^2=\frac{1}{15}\
,\ \left|\wp_{p\pm 1,d}^{p0,s}\right|^2=\frac{4}{15},
\nonumber \\
&&\left|\wp_{p1,d}^{p1,s}\right|^2=\left|\wp^{p-1,s}_{p-1,d}\right|^2=\frac{1}{45}\ ,\ 
\left|\wp_{p-1,d}^{p1,s}\right|^2=\left|\wp_{p1,d}^{p-1,s}\right|^2=\frac{2}{15},
\nonumber \\
&&\left|\wp_{p0,d}^{p\pm 1,d}\right|^2=\left|\wp^{p0,d}_{p\pm
    1,d}\right|^2=\frac{23}{75},\nonumber \\
&&\left|\wp_{p-1,d}^{p1,d}\right|^2=\left|\wp_{p1,d}^{p-1,d}\right|^2=\frac{73}{225},
\nonumber \\
&&\left|\wp_{p0,d}^{p0,d}\right|^2=\frac{34}{225}\ ,\ \left|\wp_{p1,d}^{p1,d}\right|^2=\left|\wp_{p-1,d}^{p-1,d}\right|^2=\frac{16}{75}.
\end{eqnarray}


\end{document}